# Dimensional crossover of thermal transport in few-layer graphene materials

Suchismita Ghosh[1], Wenzhong Bao[2], Denis L. Nika[1,3], Samia Subrina[1], Evghenii P. Pokatilov[1,3], Chun Ning Lau[2] and Alexander A. Balandin[1,*]

[1]*Nano-Device Laboratory, Department of Electrical Engineering and Materials Science and Engineering Program, University of California – Riverside, Riverside, California 92521 USA*

[2]*Department of Physics and Astronomy, University of California – Riverside, Riverside, California 92521 USA*

[3]*Department of Theoretical Physics, State University of Moldova, Chisinau, Republic of Moldova*

**Graphene [1], in addition to its unique electronic [2-3] and optical properties [4], revealed unusually high thermal conductivity [5-6]. The fact that thermal conductivity of large enough graphene sheets should be higher than that of basal planes of bulk graphite was predicted theoretically by Klemens [7-8]. However, the exact mechanisms behind drastic alteration of material's intrinsic ability to conduct heat as its dimensionality changes from 2-D to 3-D remain elusive. Recent availability of high-quality few-layer graphene (FLG) materials allowed us to study dimensional crossover experimentally. Here we show that the room-temperature thermal conductivity changes from K~3000 W/mK to 1500 W/mK as the number of atomic plains in FLG increases from 2 to 4. We explained the observed evolution from 2-D to bulk by the cross-plane coupling of the low-energy phonons and corresponding changes in the phonon Umklapp scattering. The obtained results shed light on heat conduction in low-dimensional materials and may open up FLG applications in thermal management of nanoelectronics.**

One of the unresolved fundamental science problems, with enormous practical implications, is heat conduction in low dimensional materials. The question of what happens with thermal conductivity when one goes to *strictly* two-dimensional (2-D) and one-dimensional (1-D) materials has attracted considerable attention [9]. Thermal transport in solids is described by thermal conductivity $K$ through Fourier's law, which states that $\boldsymbol{J}_Q = -K\nabla T$, where the heat

S. Ghosh, W. Bao, D.L. Nika, S. Subrina, E.P. Pokatilov, C.N. Lau and A.A. Balandin, UC-Riverside, 2009

flux $J_Q$ is the amount of heat transported through the unit surface per unit time and $T$ is the local temperature. This definition has been used for bulk materials as well as nanostructures. Many recent theoretical studies [9-13] suggest an emerging consensus that the *intrinsic* thermal conductivity of 2-D and 1-D anharmonic crystals is anomalous and reveals divergence with the size of the system defined either by the number of atoms $N$ or linear dimension $L$. In 2-D, $K$ diverges as $ln(N)$ while in 1-D, $K \propto N^{\alpha}$, where $0 < \alpha < 1$ (the exact value of $\alpha$ is still a subject of debate [9]). This divergence is universal property. It should not be confused with the length dependence of thermal conductivity in the *ballistic* transport regime frequently observed at low temperatures when $L$ is smaller than the phonon mean free path (MFP). Ballistic transport has been studied extensively in carbon nanotubes [14-16]. Here we focus on the diffusive transport and thermal conductivity limited by the intrinsic phonon interactions via Umklapp processes [17].

Experimental investigation of heat conduction in *strictly* 1-D or 2-D materials has essentially been absent due to lack of proper material systems. Thermal transport in conventional thin films or nanowires still retains "bulk" features because the cross-sections of these structures are measured in many atomic layers. Heat conduction in such nanostructures is dominated by the *extrinsic* effects, e.g. phonon – boundary or phonon – defect scattering [18]. The situation has changed with emergence of the mechanically exfoliated graphene – an ultimate 2-D system. It has been established experimentally that thermal conductivity of suspended single-layer graphene (SLG) is in the range $K \approx 3000 – 5000$ W/mK near room-temperature (RT) [5-6], which is clearly above the bulk graphite limit $K \approx 2000$ W/mK [7-8]. The upper bound $K$ for graphene was obtained for the largest SLG flakes examined (~20 μm x 5 μm). The extraordinary high value of K for SLG is related to the logarithmic divergence of the 2-D *intrinsic* thermal conductivity discussed above. MFP of the low-energy phonons in graphene and, correspondingly, their contribution to thermal conductivity are limited by the size of a graphene flake rather than by Umklapp scattering [7-8, 19].

The dimensional crossover of thermal transport, i.e., evolution of heat conduction as one goes from 2-D graphene to 3-D bulk, is of great interest for both fundamental science and practical applications. We addressed this problem experimentally measuring thermal conductivity of few-layer graphene (FLG) as the number of atomic planes changes from $n=2$ to $n \approx 10$. A high-quality large-area FLG sheet, where the thermal transport is diffusive and limited by the intrinsic rather than extrinsic effects, is an ideal system for such study. A number of FLG

S. Ghosh, W. Bao, D.L. Nika, S. Subrina, E.P. Pokatilov, C.N. Lau and A.A. Balandin, UC-Riverside, 2009

samples was prepared by the standard mechanical exfoliation of bulk graphite [1] and suspended across trenches in Si/SiO$_2$ wafers (Fig. 1a-c). The depth of the trenches was ~300 nm while the trench width varied in the range 1 – 5 μm. Metal heat sinks were fabricated at distances 1 – 5 μm from trench edges by the shadow mask evaporation. We utilized our previous experience in graphene device fabrication [20-21] to obtain the highest quality set of samples. The width of the suspended flakes varied from $W \approx$ 5 to 16 μm. The number of atomic layers in graphene flakes was determined with micro-Raman spectroscopy (InVia, Renishaw) through well-established decomposition of *2D* (also referred to as *G'*) band in graphene's spectra [22]. The measurements of the thermal conductivity were performed using a steady-state non-contact optical technique developed by us on the basis of micro-Raman spectroscopy [5]. The shift of the temperature-sensitive Raman *G* peak in graphene spectrum [23] defined the local temperature rise in the suspended portion of FLG in response to heating by an excitation laser in the middle of the suspended portion of the flakes (Fig. 1d). Thermal conductivity was extracted from the power dissipated in FLG, resulting temperature rise and flake geometry through the finite-element method solution of the heat diffusion equation (Fig. 1e; Methods Summary).

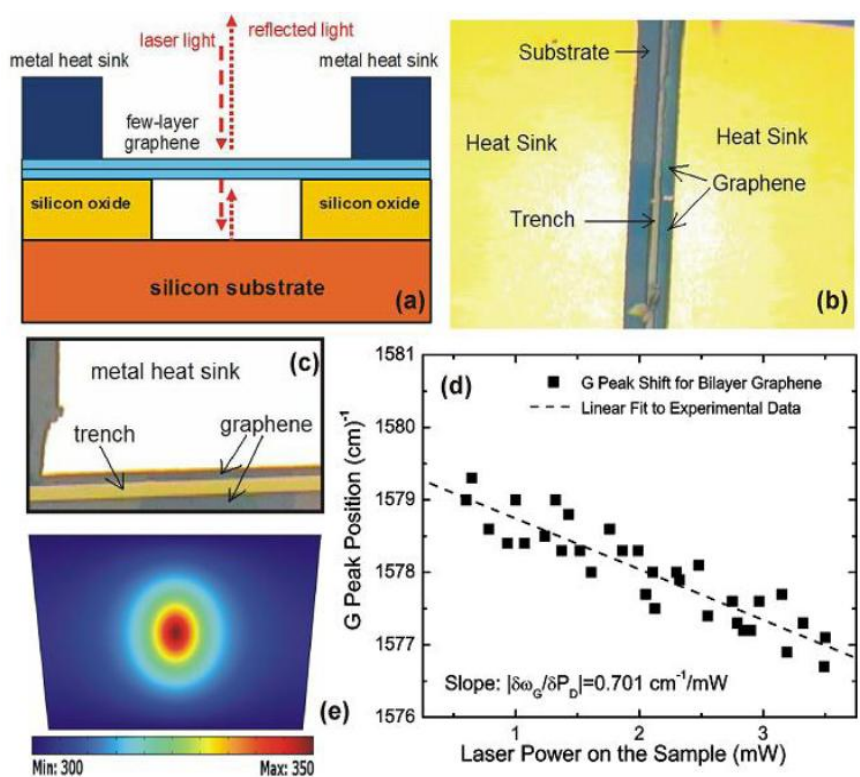

**Figure 1: Samples and measurement procedure**. a, Schematic of the thermal conductivity measurement showing suspended FLG flakes and excitation laser light. b, c, Optical microscopy images of FLG attached to metal heat sinks. d, Experimental data for Raman *G* peak position as a function of laser power, which determines the local temperature rise in response to the dissipated power. e, Finite-element simulation of temperature distribution in the flake with the given geometry used to extract the thermal conductivity.

S. Ghosh, W. Bao, D.L. Nika, S. Subrina, E.P. Pokatilov, C.N. Lau and A.A. Balandin, UC-Riverside, 2009

The power dissipated in FLG was determined through the calibration procedure based on comparison of the integrated Raman intensity of FLG's $G$ peak $\overline{I^G_{FLG}}$ and that of reference bulk graphite $\overline{I^G_{BULK}}$. Figure 2a shows measured data for FLG with $n$=2, 3, 4, ~8 and reference graphite. An addition of each atomic plane leads to $\overline{I^G_{FLG}}$ increase and convergence with the graphite while the ratio $\varsigma = \overline{I^G_{FLG}}/\overline{I^G_{BULK}}$ stays approximately independent of excitation power indicating proper calibration. In Figure 2b we present measured thermal conductivity as function of the number of atomic planes $n$ in FLG. The maximum and average $K$ values for SLG are also shown. Since thermal conductivity of graphene depends on the width of the flakes [19, 24] the data for FLG is normalized to the width $W$=5 μm to allow for direct comparison. At fixed $W$ the changes in $K$ value with $n$ are mostly due to modification of the three-phonon Umklapp scattering. The thermal transport in our experiment is in diffusive regime since $L$ is larger than the phonon MFP in graphene, which was measured [6] and calculated [25] to be around ~800 nm near RT.

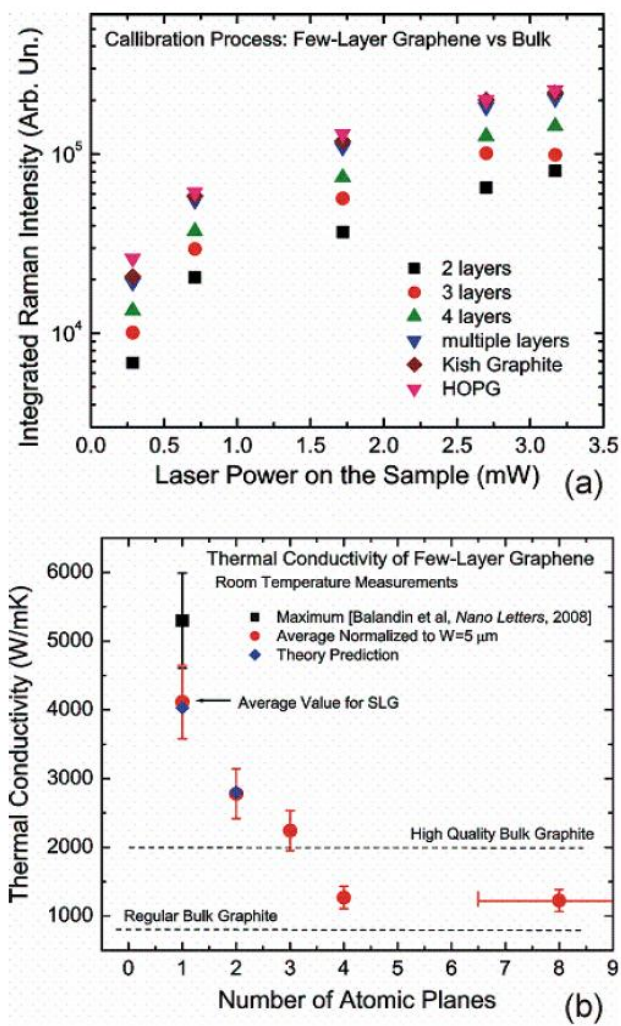

**Figure 2: Experimental data.** a, Integrated Raman intensity of $G$ peak as a function of the laser power at the sample surface for FLG and reference bulk graphite (Kish and HOPG). The data was used to determin the fraction of power absorbed by the flakes. b, Measured thermal conductivity as a function of the number of atomic planes in FLG. Dashed straigh lines indicate the range of bulk graphite thermal conductivities. Two diamond points were obtained from the first-principle theory of thermal conductivity in SLG and BLG.

S. Ghosh, W. Bao, D.L. Nika, S. Subrina, E.P. Pokatilov, C.N. Lau and A.A. Balandin, UC-Riverside, 2009

Thus, we explicitly observed heat conduction crossover from 2-D graphene to three-dimensional (3-D) graphite as the number of atomic planes changes from $n=2$ to $n=4$. The bulk value of $K$ along the basal planes is recovered at $n\approx 8$. The ambiguity of the data point for $n\approx 8$ is explained by the fact that $n$ in FLG can be determined accurately from Raman spectrum only if $n\leq 6$ [22-23].

It is striking that the measured $K$ dependence on FLG thickness $h\times n$ ($h=0.35$ nm) is opposite from what is observed for conventional thin films with the thicknesses in the range from few nm to μm ranges. In conventional films with the thickness $H$ smaller than the phonon MFP, thermal transport is dominated by the phonon – rough boundary scattering. The thermal conductivity can be estimated from as $K=(1/3)C_V\upsilon^2\tau$, where $C_V$ is the specific heat, $\upsilon$ and $\tau$ are the average phonon velocity and lifetime. When the phonon lifetime is limited by the boundary scattering, $\tau=\tau_B$, one can write, following Ziman, $(1/\tau_B)=(\upsilon/H)((p-1)/(p+1))$, which shows that $K$ scales down with the decreasing thickness (here $p$ is the specularity parameter defined by the surface roughness). In FLG with 2 or 3 atomic layer thickness there is essentially no scattering from the top and back surfaces (only the edge scattering is present [24]). Indeed, FLG is too thin for any cross-plane velocity component and for random thickness fluctuations, i.e. $p\approx 1$ for FLG. To understand the thermal crossover one needs to examine the changes in the intrinsic mechanisms limiting $K$: Umklapp scattering.

The crystal structure of BLG consists of two atomic planes of graphene bound by weak van der Waals forces resulting in the phonon dispersion different from that in SLG (Fig. 3a-b). By calculating the phonon dispersion in FLG from the first-principles [26] we were able to study phonon dynamics (Supplementary Movie) and the role of separate phonon modes. Figure 3c shows the contributions $\Delta K(q_i)$ to thermal conductivity of the phonons with the wave vectors from the interval ($q_i$, $q_{i+1}$). These contributions make up the total thermal conductivity $K=\int_0^{q_{max}}K(q)dq=\sum_1^m\Delta K(q_i)$, which is, in our calculation, limited by the phonon Umklapp and edge boundary scattering. In BLG the number of available phonon branches doubles (Fig. 3a). But the new conduction channels do not transmit heat effectively because the group velocity of $LA_2$ and $TA_2$ branches is close to zero. From the other side, increase in the thickness $h\times n$ leads to a corresponding decrease in the flux density: see region I of small $q$ where thermal transport is mostly limited by edge scattering (Fig. 3c). In the region III of

S. Ghosh, W. Bao, D.L. Nika, S. Subrina, E.P. Pokatilov, C.N. Lau and A.A. Balandin, UC-Riverside, 2009

large $q$, the number of phonon states available for the three-phonon Umklapp scattering in BLG increases by a factor of four as compared to SLG (Fig. 3d). As a result, despite the increase in the number of conduction channels in BLG, its thermal conductivity decreases because the $q$-phase-space available for Umklapp scattering increases even more. One can say that in SLG phonon Umklapp scattering is quenched and the heat transport is only limited by the edge (in-plane) boundary scattering. This is in agreement with Klemens' explanation of higher thermal conductivity of graphene compared to graphite [7-8]. It is rooted in the fundamental properties of 2-D systems discussed above. Our theory also explains molecular-dynamics simulations for "unrolled" carbon nanotubes, which revealed much higher $K$ for graphene then for graphite [27]. Two theoretical data points for SLG and BLG shown in Fig 2b are in excellent agreement with the experiment. They were obtained by accounting for all allowed Unklapp processed through extension of the diagram technique developed by us for SLG [22] to BLG: the difference between these two cases in the number of phonon modes and their energies.

Thus, we have experimentally demonstrated 2-D→3-D dimensional crossover of heat conduction in FLG. We related the increased thermal conductivity of graphene to the fundamental properties of 2-D systems and elucidated the physical mechanisms behind the evolution of heat conduction in few-atomic-thick crystals. The obtained results are important for the proposed graphene and FLG applications in lateral heat spreaders for nanoelecrtonics.

**Methods Summary**

The trenches in Si/SiO$_2$ wafers were made with the reactive ion etching (RIE, STS). The evaporated metal heat sinks ensured proper thermal contact with the flakes and constant temperature during the measurements (Fig 1b-c). FLG samples were heated with 488 nm laser (Argon Ion) in the middle of the suspended part (Fig 1a). The size of the laser spot was determined to be 0.5 – 1 μm. The diameter of the strongly heated region on the flake was somewhat larger due to the indirect nature of energy transfer from light to phonons. The laser light deposits its energy to electrons, which propagate with the Fermi velocity $v_F \approx 10^6$ m/s. The characteristic time for energy transfer from electrons to phonons is on the order of $\tau_{ph} \sim 10^{-12}$ s [28]. The measurement time in our steady-state technique is a few minutes, which is much larger than $\tau_{ph}$ but at the same time, much smaller than the time required to introduce any laser-induced damage to the sample. The correction to the hot-spot size is estimated as $l_{e\text{-}ph} = v_F \tau_{ph}$ ~ 1 μm.

S. Ghosh, W. Bao, D.L. Nika, S. Subrina, E.P. Pokatilov, C.N. Lau and A.A. Balandin, UC-Riverside, 2009

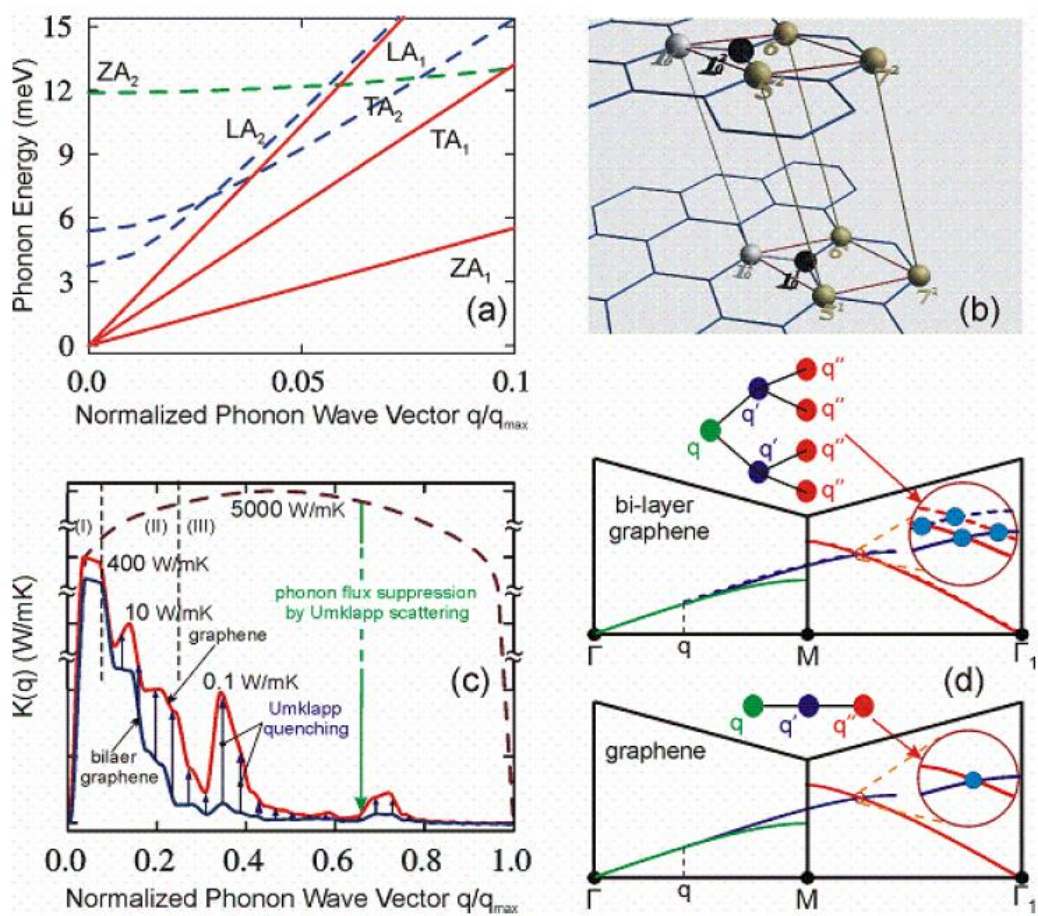

**Figure 3: Theoretical interpretation**. a, Phonon dispersion in graphene (solid curves) and bilayer graphene (added dashed curves). Note that the longitudinal acoustic ($LA_2$) and transverse acoustic ($TA_2$) phonon branches in bilayer graphene have very small slope, which translates to low phonon group velocity. b, Crystal structure of bilayer graphene. c, Contributions to thermal conductivity of different phonons indicating the range of the phonon wave vectors where Umklapp scattering is the main thermal transport limiting mechanism. d, diagram of the three-phonon Umklapp scattering in graphene and bilayer graphene, which shows that in bilayer graphene there are more states available for scattering.

Strain, stress and surface charges change the *G* peak position in Raman spectra and may lead to its splitting [29-30]. Strain and doping may also affect the thermal conductivity. To minimize these effects we selected samples where *G* peak was in its "standard" location and shape [22-23, 29-30] characteristic for the undoped unstrained graphene. No bias was applied to avoid charge accumulation [30]. It was not possible to mechanically exfoliate FLG flakes with different number of atomic plane *n* and the same geometry. To avoid damage to graphene the obtained flakes were not cut to the same shape. Instead, we solved the heat diffusion equation numerically for each sample shape to extract its thermal conductivity (Fig 1e). This was accomplished through an iteration procedure (Supplemental

Information) for the Gaussian distributed laser intensity and an effective spot size corrected to account for $l_{e\text{-}ph}$. The errors associated with the laser spot size and intensity variation were ~8%, i.e. smaller than the error associated with the local temperature measurement by Raman spectrometers (~10-13%).

**Supplemental Information** is linked to the online version of the paper at
www.nature.com/nature

S. Ghosh, W. Bao, D.L. Nika, S. Subrina, E.P. Pokatilov, C.N. Lau and A.A. Balandin, UC-Riverside, 2009

**Acknowledgements** A.A.B. acknowledges the support from DARPA – SRC through the FCRP Center on Functional Engineered Nano Architectonics (FENA) and Interconnect Focus Center (IFC) as well as from US AFOSR through contract A9550-08-1-0100.

**Author Contributions** A.A.B. conceived the experiment, led the data analysis, proposed theoretical interpretation and wrote the manuscript; S.G. performed Raman measurements; S.S. carried out finite element modelling for thermal data extraction; W.B. prepared majority of samples; C.N.L. supervised the sample fabrication; D.L.N. and E.P.P. assisted with the theory development and numerical modelling.

**Author Information** The authors declare no competing financial interests. Reprints and permissions information is available at www.nature.com/reprints. Correspondence and requests for materials should be addressed to A.A.B. (balandin@ee.ucr.edu).